\documentclass{PoS}

\title{Pion freeze-out through HBT correlation in HICs from AGS/FAIR to RHIC energies}

\ShortTitle{Pion freeze-out from AGS to RHIC}

\author{\speaker{Qingfeng Li}\thanks{In collaboration with Marcus Bleicher and Horst St\"ocker.}\\
        Frankfurt Institute for Advanced Studies (FIAS), Frankfurt\\
        E-mail: \email{liqf@fias.uni-frankfurt.de; liqfer@gmail.com}}

\abstract{In this talk we present the results of two-pion HBT
correlation at freeze-out in heavy ion collisions (HICs) from AGS to
RHIC energies. The UrQMD hadron-string transport model as well as
the CRAB analyzing program are adopted. Based on the cascade mode,
in general, the calculations are satisfying and well in line with
the experimental data although discrepancies are not negligible.
Such as: I), the HBT time-related puzzle exists at all energies.
II), at low AGS energies, the calculated volume as well as the mean
free path of pion source at freeze-out are lower than data. It
implies that a better description of interactions of particles at
early stage of HICs is required.}

\FullConference{Critical Point and Onset of Deconfinement
          4th International Workshop\\
         July 9-13 2007\\
         GSI Darmstadt,Germany}

\begin{document}

\section{Introduction}
For discovering the theoretically predicted quark gluon plasma (QGP)
the heavy ions have been collided with nucleon-nucleon
center-of-mass energies from less than $\sqrt s \sim 2.5$~GeV
(SIS/FAIR energy regime), $2.5- 20$~GeV (AGS/FAIR and SPS) up to $20
- 200$ GeV (RHIC). Indeed, there are some signals - such as
charmonium suppression, relative strangeness enhancement, etc. - of
the (phase) transition to the deconfined phase have been observed in
heavy ion collisions (HICs) at SPS energies
\cite{Matsui:1986dk,Soff:1999et,Dumitru:2001xa,Heinz:2006ur}.
Additional information about the space-time structure of the
particle emission source (the region of homogeneity) can be
extracted by Femtoscopy \cite{Lisa:2005dd} or namely
Hanbury-Brown-Twiss interferometry (HBT) \cite{Bauer:1993wq}. It is
supposed that non-trivial structures in the excitation function of
HBT quantities should be present at the energy threshold for the
onset of QGP formation \cite{Rischke:1996em}. Unfortunately, so far
the excitation functions of the HBT quantities have not shown any
{\it obvious} discontinuities within the large span of explored beam
energies
\cite{Lisa:2005dd,Antinori:2001yi,Lisa:2000hw,Ahle:2002mi,Kniege:2004pt,Kniege:2006in,
Adamova:2002wi,Adler:2004rq,Back:2004ug,Adler:2001zd,Adams:2004yc,Adcox:2002uc,
Antinori:2007jr}.

A comprehensive theoretical investigation on the excitation function
of the HBT parameters is thus highly required but still absent so
far \cite{Lisa:2005dd}. Recently, based on the UrQMD hadron-string
transport model \cite{Bass98,Bleicher99,Bratkovskaya:2004kv} and the
CRAB \cite{Pratt:1994uf,Pratthome} analyzing program, we
investigated the beam energy, transverse momentum, system-size,
centrality, and rapidity dependence of the HBT parameters $R_L$,
$R_O$, $R_S$ (dubbed as HBT radii or Pratt radii), and the cross
term $R_{OL}$ of pion source
\cite{lqf20063,Li:2007im,lqf20062,lqf2006}. In general, the
calculations are satisfying and well in line with the experimental
data although discrepancies are not negligible. Such as, I), the
calculated $R_L$ and $R_S$ values for Au+Au collisions at low AGS
energies are visibly smaller than the data if the default UrQMD
version 2.2 (cascade mode) is adopted. II), the HBT-'puzzle' with
respect to the 'duration time' of the pion source, is present at all
energies.

In this presentation, we show the beam energy and transverse
momentum dependence of the pion HBT radii, as well as some other
radius-related quantities such as the "duration-time" related
quantity $\sqrt{R_O^{2}-R_S^{2}}$ (and the $R_O/R_S$ ratio), the
freeze-out volume $V_f$, and the mean free path $\lambda_f$ of pions
at freeze-out. The standard UrQMD v2.2 in cascade mode is employed
firstly to serve as a benchmark, then we find that the HBT
time-related puzzle can be better understood with the consideration
of a potential interaction.
\section{UrQMD transport model}

The UrQMD model is based on analogous principles as the Quantum
Molecular Dynamics (QMD) \cite{Aichelin:1986wa,Aichelin:1991xy} and
the RQMD \cite{Sorge:1989dy} transport models. Similar to QMD,
hadrons are represented by Gaussian wave packets in phase space, and
the phase space of hadron $i$ is propagated according to Hamilton's
equation of motion: ${\bf \dot{r}}_i=\frac{\partial H}{\partial {\bf
p}_i}$ and ${\bf \dot{p}}_i=-\frac{\partial H}{\partial {\bf r}_i}$.
Here ${\bf {r}}$ and ${\bf {p}}$ are the coordinate and momentum of
hadron $i$. The Hamiltonian $H$ consists of the kinetic energy $T$
and the effective two-body interaction potential energy $U$. In the
standard version of UrQMD model \cite{Bass98, Bleicher99}, the
potential energies include the two-body and three-body (which can be
approximately written in the form of two-body interaction) Skyrme-
(also called as the density dependent terms), Yukawa-, Coulomb-, and
(optional) Pauli-terms as a base. Recently, in order to be more
successfully applied in the intermediate energy region ($E_b
\lesssim 2$A GeV), more potentials are considered into the model
\cite{Li:2005gf}, those are, the density-dependent symmetry
potential (essential for isospin-asymmetric reactions at
intermediate and low energies) and the phenomenologically
momentum-dependent interaction. We have found that the experimental
pion and proton directed and elliptic flows from HICs with beam
energies from $\sim 100$A MeV to $~2$A GeV can be well described
with the potentials \cite{Petersen:2006vm}. At higher energies,
i.e., AGS/FAIR, the Yukawa-, Pauli-, and symmetry- potentials of
baryons become negligible, while the Skyrme- and the
momentum-dependent parts of potentials still influence the whole
dynamical process of HICs. For example, in Ref.~\cite{Isse:2005nk},
with the help of a mean field from RQMD/S \cite{Maruyama:1996rn} and
a Jet AA Microscopic Transportation Model (JAM) it has been found
that the momentum dependence in the nuclear mean field is important
for the understanding of the proton collective flows at AGS and SPS
energies.

In the latter part of the presentation, we will also show the
importance of the mean field on the HBT time related puzzle. We
adopt the soft equation of state (EoS) with momentum dependence
(SM-EoS) which is same as that in Ref.~\cite{Isse:2005nk}.
Furthermore, as in \cite{Isse:2005nk}, the relativistic effects on
the relative distance ${\bf r}_{ij}={\bf r}_i-{\bf r}_j$ and and the
relative momentum ${\bf p}_{ij}={\bf p}_i-{\bf p}_j$ (Lorentz
transformation) employed in the two-body potentials are considered:

\begin{equation}
{\bf \tilde{r}_{ij}^2}={\bf r}_{ij}^2+\gamma_{ij}^2({\bf
r}_{ij}\cdot {\bf \beta}_{ij})^2; \label{rboost}
\end{equation}
\begin{equation}
{\bf \tilde{p}_{ij}^2}={\bf
p}_{ij}^2-(E_i-E_j)^2+\gamma_{ij}^2(\frac{m_i^2-m_j^2}{E_i+E_j})^2.
\label{pboost}
\end{equation}
In Eqs.~\ref{rboost} and \ref{pboost} the velocity-factor ${\bf
\beta}_{ij}$ and the corresponding $\gamma$-factor of $i$ and $j$
particles are defined as

\begin{equation}
{\bf \beta}_{ij}=\frac{{\bf p}_i+{\bf p}_j}{E_i+E_j}, \mbox{  and  }
\gamma_{ij}=\frac{1}{\sqrt{1-{\bf \beta}_{ij}^2}}.
\end{equation}

Similar to RQMD, the collision term of the UrQMD model treats 55
different baryon species (including nucleon-, delta- and hyperon-
resonances with masses up to 2.25 GeV) and 32 different meson
species, including (strange) meson resonances with masses up to 2.0
GeV as tabulated in the PDG \cite{PDG2000}, as well as the
corresponding anti-particles, i.e. full baryon/antibaryon symmetry
is included. Isospin is explicitly treated as well. For hadronic
continuum excitations a string model is used. Starting from the
version 2.0, the PYTHIA is incorporated into UrQMD in order to
investigate the jet production and fragmentation at RHIC energies
\cite{Bratkovskaya:2004kv}.
\section{Analyzing process}
To calculate the two-particle correlator, the CRAB program is based
on the formula:
\begin{equation}
C({\bf k},{\bf q}) = \frac {\int d^4x_1 d^4x_2 g(x_1,{\bf p_1})
g(x_2,{\bf p_2}) |\phi({\bf q}, {\bf r})|^2} {{\int d^4x_1
g(x_1,{\bf p_1})} {\int d^4x_2 g(x_2,{\bf p_2})}}. \label{cpq}
\end{equation}
Here $g(x,{\bf p})$ is the probability for emitting a particle with
momentum ${\bf p}$ from the space-time point $x = ({\bf r}, t)$.
$\phi({\bf q}, {\bf r})$ is the relative two-particle wave function
with ${\bf r}$ being their relative position.
$\bf{q}=\bf{p}_2-\bf{p}_1$ and
$\textbf{k}=(\textbf{p}_{1}+\textbf{p}_{2})/2$ are the relative
momentum and the average momentum of the two particles. Due to the
underlying quantum statistics, this correlator can be fitted
approximately by a Gaussian form. Using Pratt's three-dimensional
convention (the LCMS system), the correlation function in Gaussian
form reads
\begin{equation}
C(q_O,q_S,q_L)=1+\lambda {\rm
exp}(-R_L^2q_L^2-R_O^2q_O^2-R_S^2q_S^2-2R_{OL}^2q_Oq_L).
\label{fit1}
\end{equation}
Here $q_i$ and $R_i$ are the components of the pair momentum
difference $\bf{q}$ and the homogeneity length (HBT radii) in the
$i$ direction, respectively. The pre-factor $\lambda$ is the
incoherence parameter and lies between 0 (complete coherence) and 1
(complete incoherence) in realistic HICs. The term $R_{OL}^{2}$ is
called cross-term and vanishes at mid-rapidity for symmetric
systems, while it deviates from zero at large rapidities
\cite{Chapman:1994yv,lqf20062}.

We compare our calculations of the Pratt parameters (at midrapidity)
of the pion source with experimental data for the following central
collisions of heavy nuclei:
\begin{enumerate}
\item
Au+Au at the AGS beam energies $E_b=2$, $4$, $6$, and $8$A GeV
($<11\%$ of the total cross section $\sigma_T$), a rapidity cut
$|Y_{cm}|<0.5$ ($Y_{cm}=\frac{1}{2}{\rm
log}(\frac{E_{cm}+p_\parallel}{E_{cm}-p_\parallel})$, $E_{cm}$ and
$p_\parallel$ are the energy and longitudinal momentum of the pion
meson in the center-of-mass system) is employed. The experimental
(E895) data are taken from \cite{Lisa:2000hw}.

\item
Au+Au at the AGS beam energy $11.6$A GeV (the $<5\%$ most central
collisions), a rapidity cut $|Y_{cm}|<0.5$ is employed. The
experimental (E802) data are taken from \cite{Ahle:2002mi}.

\item
Pb+Pb at the SPS beam energies $E_b=20$, $30$, $40$, $80$, and
$160$A GeV ($<7.2\%\sigma_T$ of most central collisions), a
pion-pair rapidity cut $|Y_{\pi\pi}|<0.5$
($Y_{\pi\pi}=\frac{1}{2}{\rm log}(\frac{E_1+E_2+p_{\parallel
1}+p_{\parallel 2}}{E_1+E_2-p_{\parallel 1}-p_{\parallel 2}})$ is
the pair rapidity with pion energies $E_1$ and $E_2$ and
longitudinal momenta $p_{\parallel 1}$ and $p_{\parallel 2}$ in the
center of mass system) is employed. The experimental (NA49) data are
taken from \cite{Kniege:2004pt,Kniege:2006in}.

\item
Pb+Au at the SPS beam energies $E_b=40$, $80$, and $160$A GeV (the
$<5\%$ most central collisions), the pion-pair rapidity cut
$Y_{\pi\pi}=-0.25\sim 0.25$, $-0.5\sim 0$, and $-1.0\sim -0.5$ are
chosen. The experimental (CERES) data are taken from
\cite{Adamova:2002wi}.

\item
Au+Au at the RHIC energies $\sqrt{s_{NN}}=30$ ($<15\%\sigma_T$),
$62.4$ ($<15\%\sigma_T$), $130$ ($<10\%\sigma_T$), and $200$ GeV
($<5\%\sigma_T$). Here a pseudo-rapidity cut $|\eta_{cm}|<0.5$
($\eta_{cm}=\frac{1}{2}{\rm
log}(\frac{p+p_\parallel}{p-p_\parallel})$, ($p$ is the momentum of
the pion) is employed. The experimental (PHOBOS, STAR, and PHENIX)
data are taken from
\cite{Back:2004ug,Adler:2001zd,Adcox:2002uc,Adler:2004rq,Adams:2004yc}.
\end{enumerate}
\section{The HBT radii at AGS, SPS, and RHIC energies}
\begin{figure}
  \centering
  \begin{minipage}[c]{.55\textwidth}
    \centering
\includegraphics[angle=0,width=1\textwidth]{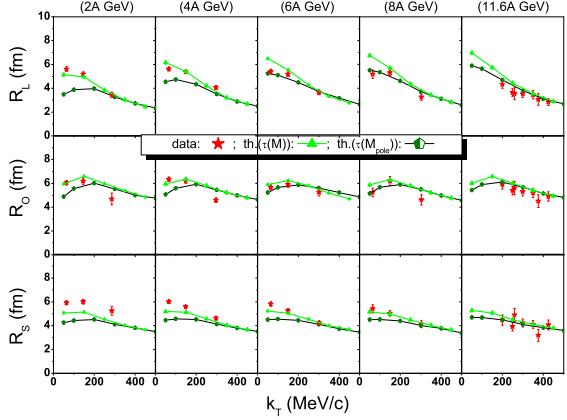}
  \end{minipage}
  \begin{minipage}[c]{.35\textwidth}
    \centering
\caption{Transverse momentum $k_T$ dependence of the HBT-radii
$R_L$, $R_O$, and $R_S$ at AGS energies. The calculated results with
and without considering the mass dependence of resonance lifetimes
are demonstrated. Experimental data are shown
\cite{Lisa:2000hw,Ahle:2002mi} with scattered stars. } \label{fig1}
\end{minipage}
\end{figure}

Fig.\ \ref{fig1} shows the $k_T$
($\textbf{k}_T=(\textbf{p}_{1T}+\textbf{p}_{2T})/2$) dependent HBT
radii $R_L$ (top), $R_O$ (middle), and $R_S$ (bottom) of the
$\pi^-\pi^-$-pair for AGS energies $2$, $4$, $6$, $8$, and $11.6$A
GeV (from left to right). The cascade mode is employed. We notice
that the radii $R_L$ and $R_S$ are somewhat smaller than the data
especially at lower beam energies, if a constant width for
resonances (lines with diamonds) is used. Fig.\ \ref{fig1} also
includes the results with the mass dependent lifetime of resonances
(lines with triangles). With this treatment, the resonances with
their small invariant masses decay later and hence the expanded
fireball becomes larger as compared to the standard (mass
independent) treatment. At large $k_T$ as well higher beam energies,
this effect is reduced. It is interesting to see that the result
with a mass-dependent treatment of resonance lifetimes can matches
the data much better in the AGS energy region. It is also seen that
this mass-dependence has almost no effect on the ratio between $R_O$
and $R_S$ values.

\begin{figure}
\centering
\begin{minipage}[t]{0.48\textwidth}
    \centering
\includegraphics[angle=0,width=1\textwidth]{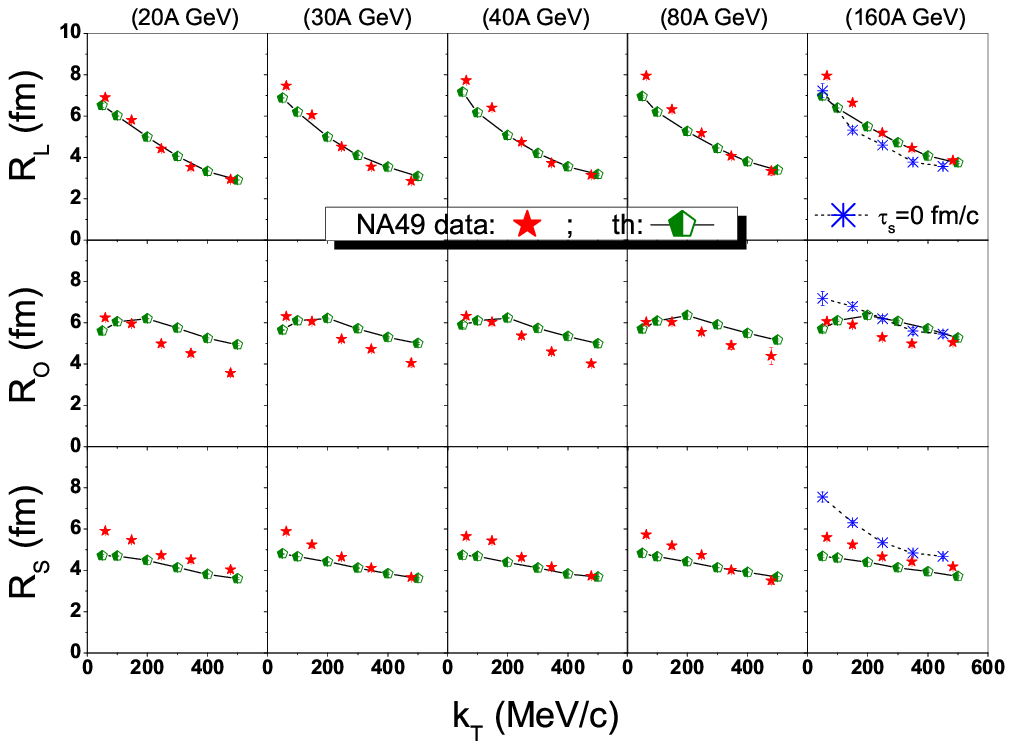}
\caption{$k_T$ dependence of the Pratt-radii at SPS energies.
Preliminary NA49 data are taken from \cite{Kniege:2006in}. At
$E_b=160$A GeV the calculation results of the HBT-radii with a
vanishing formation time for strings are also presented. }
\label{fig2}
\end{minipage}
  \begin{minipage}[t]{0.48\textwidth}
    \centering
\includegraphics[angle=0,width=1\textwidth]{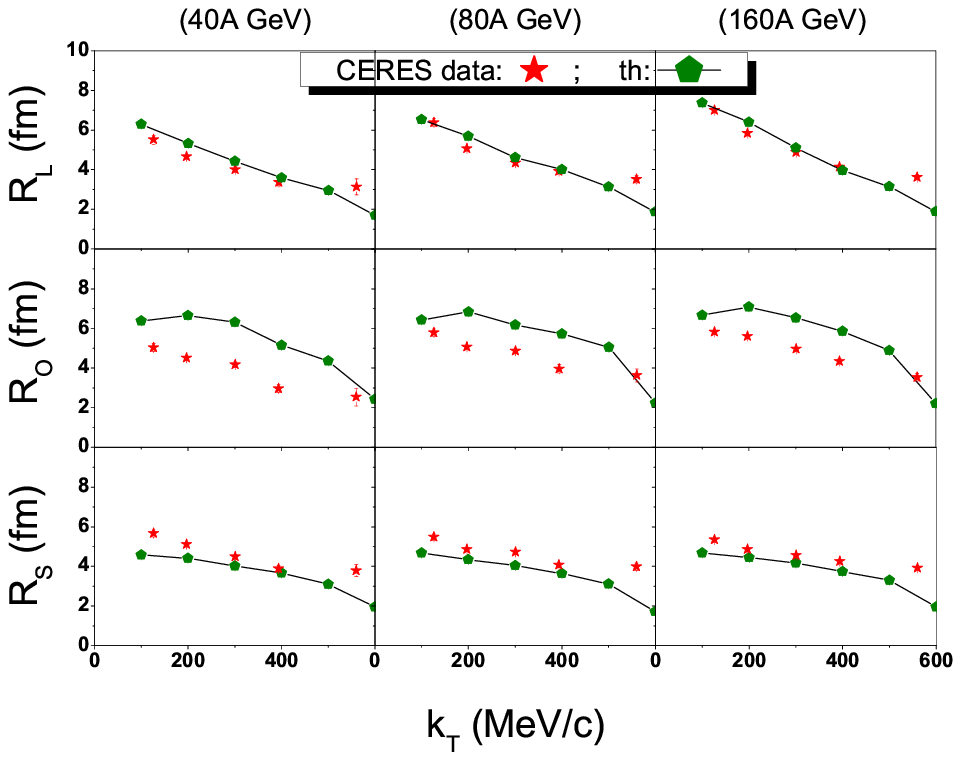}
\caption{$k_T$ dependence of the Pratt-radii at SPS energies. CERES
data are taken from \cite{Adamova:2002wi}.} \label{fig3}
 \end{minipage}%
\end{figure}

In Figs.\ \ref{fig2} and \ref{fig3} we show the $k_T$ dependence of
the HBT- radii $R_L$ (top plots), $R_O$ (middle plots), and $R_S$
(bottom plots) at SPS energies. In Fig.\ \ref{fig2} the results are
compared with preliminary NA49 data \cite{Kniege:2006in}. In Fig.\
\ref{fig3} the results are compared with CERES data
\cite{Adamova:2002wi}. Furthermore, the $\pi^{-}-\pi^{-}$
correlations are calculated in Fig.\ \ref{fig2} while the
two-charged-pion correlations (including two-$\pi^-$ and two-$\pi^+$
mesons) are calculated in Fig.\ \ref{fig3}. Firstly, it is very
interesting to see that the present calculations can reproduce the
$k_T$-dependence of HBT radii $R_L$ and $R_S$ fairly well. Only at
small $k_T$, the calculated $R_L$ and $R_S$ values are seen up to
$25\%$ lower than data. While for the $R_O$ values, the calculations
are shown larger than both NA49 and CERES data especially at
relatively large $k_T$. By comparing the NA49 data with the CERES
data for central Pb+Pb ($\sigma/\sigma_T<7.2\%$) and central Pb+Au
($\sigma/\sigma_T<5\%$) collisions, one observes that the CERES
$R_O$ data are somewhat smaller than the recent published NA49 data
\cite{Kniege:2006in} especially at large $k_T$ and low beam energy
$E_b=40$A GeV although the recent NA49 data at large $k_T$ have
already been driven down visibly when comparing with those
preliminary data in the previous publication \cite{Kniege:2004pt}.
The origin of the difference between NA49 and CERES data is still
not quite clear. At 160 A GeV (shown in Fig.\ \ref{fig2}) the HBT
radii are also calculated by adopting zero formation time for
strings ($\tau_s=0$ fm$/$c). As a result, the $k_T$-dependence of
the HBT radii becomes steeper, and the values of $R_S$ increase and
approach the calculated $R_O$ values. It should be noted that,
although a shorter formation time is apt to explain the
``HBT-puzzle", as well the elliptic flow, the absolute values of
HBT-radii are not well in line with data.

\begin{figure}
  \centering
  \begin{minipage}[c]{.55\textwidth}
    \centering
\includegraphics[angle=0,width=1\textwidth]{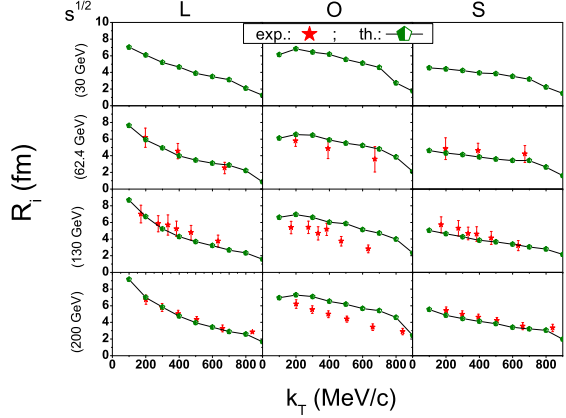}
  \end{minipage}
  \begin{minipage}[c]{.35\textwidth}
    \centering
\caption{$k_T$ dependence of the Pratt-radii $R_L$, $R_O$, and $R_S$
at RHIC energies. Experimental data are shown
\cite{Back:2004ug,Adler:2001zd,Adcox:2002uc,Adler:2004rq,Adams:2004yc}
at $\sqrt{s_{NN}}=62.4$, $130$, and $200$ GeV. } \label{fig4}
 \end{minipage}
\end{figure}

Fig.\ \ref{fig4} gives the $k_T$ dependence of the Pratt-radii $R_L$
(left plots), $R_O$ (middle plots), and $R_S$ (right plots) at RHIC
energies. Both the absolute values and the decrease of the
Pratt-radii $R_L$ and $R_S$ with transverse momentum is reproduced
by the present model calculations very well. Here, it is also seen
that the calculated $k_T$-dependence of  $R_S$ is somewhat flatter
than that of $R_L$, which implies that flow effect on the
$k_T$-dependence of the Pratt-radii is still important. Besides the
flow effect, the surface-like emission charactistic of microscopic
models should play significantly role on HBT parameters as well
because also other Cascade/Boltzmann model-calculations (see e.g.,
the RQMD \cite{Lisa:2000hw,Lisa:2005dd}, the HRM
\cite{Humanic:2005ye}, and the AMPT \cite{Lin:2002gc}) can reproduce
the $k_T$ dependence of Pratt radii (almost) equally well. The UrQMD
calculations of $R_L$ and $R_S$ reproduce the experimental data well
within the error bars, while the calculated $R_O$'s are also larger
than the experimental data --- the $R_O$  is about $25\%$ too large.

\section{Excitation function of the $V_f$, the $\lambda_f$, and the $\sqrt{R_O^{2}-R_S^{2}}$ (and $R_O/R_S$) of pions}

\begin{figure}
  \centering
  \begin{minipage}[c]{.75\textwidth}
    \centering
\includegraphics[angle=0,width=1\textwidth]{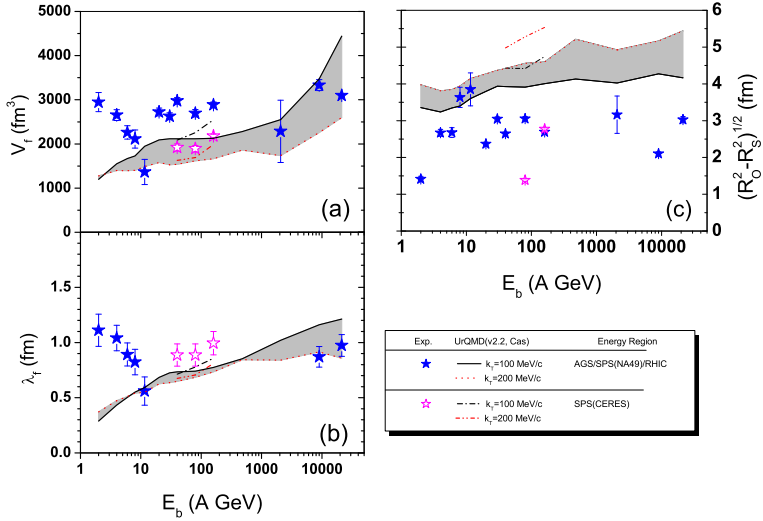}
  \end{minipage}
  \begin{minipage}[c]{.23\textwidth}
    \centering
\caption{(a): Excitation function of the $V_f$ at transverse momenta
between $k_T=100$MeV and $200$MeV (gray area), compared with data in
this $k_T$-region. (b): Excitation function of the $\lambda_f$ of
pions at freeze-out. (c): Excitation function of the duration-time
related quantity $\sqrt{R_O^{2}-R_S^{2}}$. } \label{fig5}
  \end{minipage}
\end{figure}

\begin{figure}
  \centering
  \begin{minipage}[c]{.65\textwidth}
    \centering
\includegraphics[angle=0,width=1\textwidth]{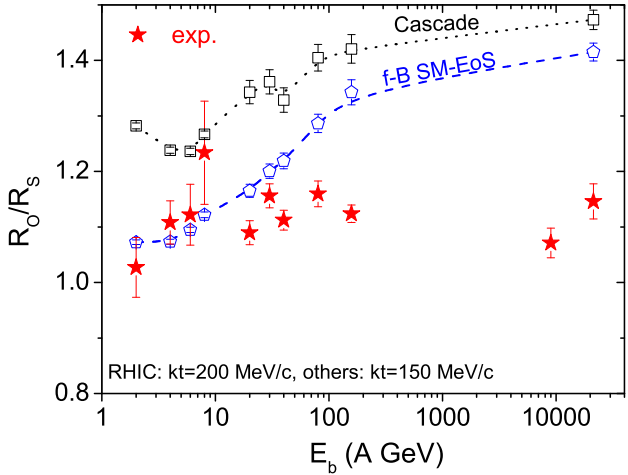}
  \end{minipage}
  \begin{minipage}[c]{.33\textwidth}
    \centering
\caption{Excitation function of the $R_O/R_S$ ratio at small $k_T$.
The data are indicated by solid stars. The dotted lines with open
rectangles are results under cascade mode, the dashed lines with
open diamonds represent the calculations with potentials of formed
baryons. } \label{fig6}
  \end{minipage}
\end{figure}

Fig.\ \ref{fig5} (a) shows the excitation function of the pion
source volume $V_f$ at freeze-out, calculated as
\cite{Adamova:2002ff} $ V_f=(2\pi)^\frac{3}{2}R_LR_S^2$. The
calculations with default cascade mode are shown at $k_T=100\pm
50$~MeV (full line) and $200\pm 50$~MeV$/c$ (dotted line),
respectively. The gray areas between the two lines are shown for
better visibility. The data are at $k_T\sim 150$MeV$/c$ for AGS-E895
and SPS-NA49, at $k_T\sim 170$MeV$/c$ for reaction at
$\sqrt{s_{NN}}=130$ GeV, at $k_T\sim 200$MeV$/c$ for reactions at
AGS-E802, SPS-CERES, and $\sqrt{s_{NN}}=62.4,200$ GeV. Since the
experimental data from NA49 \cite{Kniege:2004pt,Kniege:2006in} and
from CERES \cite{Adamova:2002wi} collaborations overlap at beam
energies $40$, $80$, and $160$A GeV, we show the calculations and
data with respect to CERES energies separately as dashed-dotted
lines and open symbols. Fig.\ \ref{fig5} (a) shows clearly that the
UrQMD cascade calculations do provide a reasonable freeze-out volume
for the pion source at RHIC energies. At SPS energies, the agreement
is fine with CERES data while it slightly underpredicts those of
NA49. Towards even lower energies, the model underpredicts the
measured freeze-out volume due to the omission of the strong
interaction potential and other in-medium effects. E.g. at $E_b=2A$
GeV, the measured $V_f$ is about $2-3$ times larger than calculated
value. As stated in Fig.\ \ref{fig1}, a mass-dependent lifetime of
resonances accounts for an improvement of the HBT-radii at small
$k_T$ and hence reproduce the data better.

The mean free path $\lambda_f$ of the pions at freeze-out is
expressed as \cite{Adamova:2002ff}
\begin{equation}
\lambda_f=\frac{V_f}{N\sigma}=\frac{V_f}{N_N\sigma_{\pi
N}+N_{\pi}\sigma_{\pi\pi}} \label{lam1}.
\end{equation}
with the averaged pion-nucleon cross section $\sigma_{N\pi}=72$~mb
and the averaged pion-pion cross section $\sigma_{\pi\pi}=13$~mb
(Note that within the present model calculations these values are
slightly energy dependent. However, here we have adopted the
explicit numbers from Ref. \cite{Adamova:2002ff} to compare to the
results presented there). The nucleon and pion multiplicities $N_N$
and $N_{\pi}$ are calculated as

\begin{equation}
N_{N,\pi}=y_{th} \cdot \sqrt{2\pi} \cdot
\left.\frac{dN_{nucleons,pions}}{dy}\right.|_{y_{mid}}
\label{NNNpi1}.
\end{equation}
using the assumption of a thermal equilibrated system at freeze-out
with a temperature $T_f=120$ MeV. Here, $y_{th}$ is the estimated
thermal homogeneity scale in rapidity at a certain $k_T$ and $T_f$,
and is given by the  expressions: $y_{th}={\rm
arctanh}(\langle\beta_{th}\rangle)$, with
$\langle\beta_{th}\rangle=\sqrt{1+\langle\gamma\rangle^{2}}/\langle\gamma\rangle$
and
$\langle\gamma\rangle=1+1/3\left(K_1(m_T/T_f)/K_2(m_T/T_f)-1\right)
+T_f/m_T$. Here $K_n(z)$ is the modified Bessel function of order
$n$ and $m_T=\sqrt{m_\pi^2+k_T^2}$. $dN/dy|_{y_{mid}}$ is the
rapidity density of pion (nucleons) at mid-rapidity. Recent
calculations using the present UrQMD transport model
\cite{Bratkovskaya:2004kv} have shown that the calculated pion and
nucleon yields are reasonably in agreement with data.

Fig.\ \ref{fig5} (b) shows the excitation function of $\lambda_f$ of
pions at freeze-out. The experimental value for $\lambda_f$ at
$\sqrt{s_{NN}}=200$GeV is obtained with the help of recent $dN/dy$
data in \cite{Adler:2003cb}, at all other energies the $\lambda_f$
data are taken from \cite{Adamova:2002ff}. It is seen that the
theoretical $\lambda_f$ value increases gradually from $\sim 0.5$ to
$\sim 1$~fm from AGS to highest RHIC energies with a weak dependence
on $k_T$. The experimental values of $\lambda_f$ are also between
$0.5-1$~fm. The observation (both experimentally and theoretically)
of a nearly energy independent mean free path on the order of
$0.7$~fm at pion freeze-out is rather surprising. Physically it has
been interpreted as a rather large opaqueness of the pion source at
break-up \cite{Heiselberg:1997vh,Tomasik:1998qt}.

The HBT duration time ''puzzle'', i.e. the fact of the theoretical
quantity $\sqrt{R_O^{2}-R_S^{2}}$ being larger than extracted from
the data, is present at all investigated energies (see Fig.\
\ref{fig5} (c)): The calculated values of $\sqrt{R_O^{2}-R_S^{2}}$
are about $3.5\sim 5$~fm while the measured ones are $1.5\sim 4$~fm.
Many efforts have been put forward over the last years to clarify
this issue
\cite{Lin:2002gc,Cramer:2004ih,Humanic:2005ye,Pratt:2005hn,Pratt:2005bt,lqf2006,lqf20062,lqf20063}.
Here, we show in Fig.\ \ref{fig6} the potential effect on the
excitation function of the $R_O/R_S$ ratio at the small $k_T$. As
seen in Fig.\ \ref{fig5} (c), the $R_O/R_S$ ratio, which is
equivalent to the quantity $\tau \sim \sqrt{R_O^{2}-R_S^{2}}$, is
larger than the experimentally observed values at all investigated
energies, if the cascade mode (dotted lines with open rectangles) is
employed. When the SM-EoS is considered for the formed baryons
(solid lines with open diamonds), the $R_O/R_S$ ratio is seen
obviously smaller than that with cascade mode and reproduces the
(energy dependence of) data at AGS energies. At SPS and RHIC,
however, the $R_O/R_S$ ratio is still increasing monotonically with
increasing beam energies and deviates from data again. At RHIC
energies, the ratio approaches the one with cascade mode. It implies
that the potential of formed baryons is increasingly losing its
importance with increasing beam energies. At SPS and RHIC energies
the deviation from data might be interpreted by the absence of the
interactions of unformed particles from string fragmentation.
Investigations are in progress.

\section{Conclusion and Outlook}
To summarize, we show the transverse momentum and beam energy
dependence of the HBT radii $R_L$, $R_O$, and $R_S$, the quantity
$\sqrt{R_O^{2}-R_S^{2}}$ (and the $R_O/R_S$ ratio), the volume
$V_f$, and the mean free path $\lambda_f$ of pions at freeze-out for
heavy systems with energies from AGS to RHIC. In general, the model
calculations with UrQMD v2.2 (cascade mode) are in line with the
data over the whole inspected energy range. We also find a nearly
constant mean free path for pions on the order of $\lambda_f=0.7$~fm
which indicating a significant opaqueness of the source.

Discrepancies especially in the lower AGS energy region are found
and have to be resolved. The HBT duration-time related "puzzle" is
present at almost all energies. The consideration of potentials for
formed and unformed particles provides new insights into the origin
of the time-related puzzle and the dynamics of HICs especially at
early stage.

\section*{Acknowledgements}
We would like to thank the Frankfurt Center for Scientific Computing
(CSC). This work is partly supported by GSI, BMBF, DFG and
Volkswagenstiftung.

\end{document}